\documentclass[square]{ws-procs11x85}

\newtheorem{thm}{Theorem}[section]

\newtheorem{lem}[thm]{Lemma}

\begin{document}
\pagestyle{empty}
\title{Towards a theoretical understanding of false positives in DNA motif finding}
\author{Amin Zia and Alan M. Moses}
\address{Department of Cell \& Systems Biology \\
University of Toronto \\
25 Willcocks Street \\
Toronto, Ontario\\ 
Canada M5S 3B2\\
amin.zia@utoronto.ca\\
alan.moses@utoronto.ca}

\begin{abstract}
Detection of false-positive motifs is one of the main causes of low performance in motif finding methods. It is generally assumed that false-positives are mostly due to algorithmic weakness of motif-finders \cite{tompa,sandve,hu}. Here, however, we derive the theoretical dependence of false positives on dataset size and find that false positives can arise as a result of large dataset size, irrespective of the algorithm used. Interestingly, the false-positive strength depends more on the number of sequences in the dataset than it does on the sequence length. As expected, false-positives can be reduced by decreasing the sequence length or by adding more sequences to the dataset. The dependence on number of sequences, however, diminishes and reaches a plateau after which adding more sequences to the dataset does not reduce the false-positive rate significantly. Based on the theoretical results presented here, we provide a number of intuitive rules of thumb that may be used to enhance motif-finding results in practice. 
\end{abstract}
\vspace{5mm}
\noindent \textbf{Introduction}\\
\noindent Because binding of sequence specific transcription factors to their recognition sites in non-coding DNA is an important step in the control of gene expression, the development of computational methods to identify transcription factor binding motifs in non-coding DNA has received much attention in computational biology. The low information content of transcription factor binding motifs implies difficulty for computational analyses.  For example, given a known binding motif, identification of bona fide examples is always plagued by false positives - the so-called Futility Theorem \cite{wasserman}. 

An even more challenging computational problem is the de novo identification of transcription factor binding motifs (so-called motif-finding), for which there are many available tools (for tutorials on different methods see \cite{das,moses} and references therein). Despite the substantial algorithm development effort in this area, recent comprehensive benchmark studies \cite{tompa,sandve,hu} revealed that the performance of DNA motif-finders leaves room for improvement in realistic scenarios, where known transcription factor binding sites have been planted in test sequence sets. 
One of the major problems is that DNA motif-finders can identify seemingly strong candidate motifs, even when randomly chosen sequences are provided as the input. This has led to simulation-based approaches to identify the bona fide motifs where the motif-finding algorithm is repeated several times on random data and the p-value is computed as the fraction of motifs with better scores than the motif identified in real data \cite{harbison}. While feasible for expert computational biologists, this approach requires significant computational resources, and is not practical for most biological users. 

We argue that part of low performance in motif finding algorithms is due to the statistical nature of large sequence datasets: when the dataset is large enough, any structure can occur by chance. We formalize this idea using information theory, to obtain a remarkably simple analytical relationship between the size of the sequence search space and the strength of the false-positive motifs. Interestingly, our analysis shows that for biologically realistic dataset sizes and motif strengths, false positives as strong as real transcription factor binding sites are quite likely to arise.  This represents an extension of the ``Futility Theorem'' \cite{wasserman} to the de novo motif-finding problem.
Results

\vspace{5mm}
\noindent \textbf{Results}\\
\noindent \textbf{\textit{Motif-finders are expected to find strong signals in random DNA sequences:}} We represent patterns in DNA sequence families (called motifs) as probability matrices, where each column specifies the distribution of the DNA letters. The underlying idea here is to quantify the probability of observing motifs of a certain strength in a set of random sequences using large-deviations theory. Suppose that the set of all motifs, $X$, is generated according to a random nucleotide background distribution $g$ (for instance, $g$ can be the genome-wide distribution of nucleotides). It is expected that all nucleotides in motifs will have frequencies close to those in $g$. Therefore, motifs that have a distribution significantly different from $g$ (i.e. the false-positives in our case) are considered as the rare events that are far from expectation. We use the large-deviations theory, in particular SanovÕs theorem \cite{cover} to measure the probability of these rare events. We, then, derive expected size of the set $X$ above which the observation of strong motifs becomes likely to be due to chance.

Let a DNA motif with $W$ columns have a distribution or probability matrix $f$ (see Fig. 1 and Methods for definition of motif finding problem parameters). The difference between the distribution of the motif, $f$, and the background distribution, $g$, is measured using the Kullback-Leibler (KL) divergence \cite{cover}, also known as the biological information content \cite{information}\cite{stormo}, defined as in the following:
\begin{equation}
D(f,g) = I_{seq}(f,g) \triangleq \sum_{j=1}^W \sum_{k \in \{T,C,A,G\}} f_{jk} \log \frac{f_{jk}}{g_k}
\end{equation} where $f_{jk}$ is the relative frequency of base $k$ in column $j$ of the motif, and $g_k$ is the background distribution of base $k$ (e.g. the genome-wide distribution of nucleotide bases). Throughout the text we use the strength of a motif and its information content, interchangeably to refer to $D(f,g)$ and Iseq. 

Our main theoretical result is as follows. Consider the ``one-occurrence-per-sequence'' motif-finding model where each of $n$ sequences is assumed to have exactly one occurrence of a motif of width $W$. The expected sequence length, $L$, in order to observe at least one motif with a probability matrix (PM) diverged from the background, $g$, by at least $D(f, g)$ is given by:
\begin{equation}
L \approx \frac{W2^{D(f, g)}}{(n+1)^{W(|{\cal A}|-1)/n}} 
\end{equation} where $|A|$ is the cardinality of the set $A$, e.g. $|A| = 4$ for DNA sequences. According to this theorem, if the length of DNA sequences is approximately (or larger than) $L$, false-positive motifs with information content $D(f,g)$ can occur by chance. Please see Appendix A for the proof of this theorem. 

Figure 2 show the expected length of sequence, $L$ as a function of motif information content, $D(f,g)$, for DNA sequences with typical motif-finding parameters and $W=5$, $W=10$ and $W=15$, respectively. Each graph illustrates $L$, at which false-positive motifs with strength $D(f,g)$ are expected to occur by chance. 

The dependency of false-positives on $n$ is stronger compared to the dependency on $L$. As an example, for motifs with $W=10$ (Fig. 2b), a threefold increase of $n$ (while keeping $L$ constant) reduces $D(f,g)$ by the same amount as if $L$ were increased by $2$ orders of magnitude (while keeping $n$ unchanged). However, the dependency of false-positives on $n$ decreases with $n$ and reaches a plateau for larger $n$ suggesting that in order to reduce the false-positive rate only a sufficient number of sequences in the dataset is necessary (Fig. 3).

Finally, the false-positive information content, $D(f,g)$, is approximately linear in $W$ in the range of interest (Fig. 4). Therefore, given a motif-strength of interest, detecting real motifs with smaller width is easier and less prone to false-positives.

\vspace{3mm}
\noindent \textbf{\textit{MEME performance is consistent with the theoretical expectations:}} To confirm our theoretical results, we conducted a set of experiments using the MEME software \cite{meme,memeweb}, because it implements the one-occurrence-per-sequence set up that we have treated theoretically (see Methods for detail of the experiment setup). We ran MEME on a set of randomly generated sequences and asked MEME to report the most significant motif. The detected motifs are therefore false-positives. The results from MEME are plotted in Fig. 2 and are consistently following the theoretical predictions. 

\vspace{3mm}
\noindent \textbf{\textit{Simple rules of thumb for DNA motif-finding:}} The theoretical predictions provide sequence lengths above which observation of motifs with given strengths or less are most probably due to chance than any biological reason. Therefore, to reduce the false-positive strength in experimental design, it is generally desired to move towards weaker motifs (using Eq. 2 or using the curves in Fig. 2). We have the following rules of thumb for this purpose:
\begin{enumerate}
\item As it is intuitively expected, it is generally preferred to use shorter sequences (when it is biologically plausible) to avoid unnecessary false-positives.
\item Adding more sequences to the dataset reduces the false-positive rate considerably (e.g. using $30$ sequences compared to $10$ reduces the false-positive motif strengths by more than $6$ bits ($\%25$) for $W=10$, see Fig. 3). This effect, however, diminished for larger $n$ (e.g. increasing $n$ from $30$ to $50$ has only $2$ bits reduction in motif strengths, see Fig. 3). This suggests that in order to reduce false-positive rate in ``one-occurrence-per-sequence'' motif finding, only a ``sufficient'' number of sequences is needed.
\item The dependency of false-positives (the strength of false-positive motifs) on $L$ is weaker than dependency on $n$. Therefore, using many sequences (but not too many) is generally preferred to using shorter sequences. 
\item Given n sequences of length $L$ and a width $W$ for potential motifs, Eq. 2 gives expected strength of false-positive motifs. Detected motifs that do not greatly exceed this expected strength should be doubted, while motifs that are stronger than the expected value are most probably not false-positives. 
\item Given a certain strength of interest, detection of motifs with smaller width is less prone to false-positives and therefore easier.
\end{enumerate}

\vspace{3mm}
\noindent \textbf{\textit{Examples of applications:}} In using the theoretical results in Eq. 2 or the graphs in Fig. 2, it is generally desired to move towards weaker motifs (towards the left on the graphs). To illustrate this we chose the ZFP423 and the TATA-box motifs from the Jaspar database \cite{jaspar} with $D(f,g)= 17.93$  and $D(f,g)=10.20$, respectively. We show that is it difficult to detect ZFP423 in sequences of length $1000$, but it can be detected in shorter sequences (Fig. 5). Similarly, we show that it is very difficult to detect the TATA-box using $20$ sequences, but it is possible if this is increased to $30$ or if the motif is trimmed to include only the core positions (Fig. 5). 

\vspace{5mm}
\noindent \textbf{Discussion}\\
\noindent \textbf{\textit{Application to protein sequences:}} The theoretical analysis here can be applied directly for motif-finding in sequences of different alphabets. In particular, the proposed equations can be used for protein sequences by replacing $|A|=4$ with $|A|=20$ corresponding to $20$ amino-acid residues. It is easy to verify in Eq. 2, that by this modification, i.e. changing $|A|=4$ to $|A|=20$, the expected length, $L$, increases exponentially. This suggests that, under equivalent settings, the false-positive rate in the protein motif finding is exponentially lower than in the DNA motif finding. 

\vspace{3mm}
\noindent \textbf{\textit{Extension to other motif-finding models:}} The proposed method here assumes the ``one-occurrence-per-sequence'' model in motif finding (similar to the OOPS model in MEME \cite{memeweb}). However, the analysis is extendable to other models by appropriately redefining the space of all motifs in the dataset. See Appendix B for extension of Eq. 2 to the cases where each sequence can carry either zero or one motif (similar to ZOOPS model in MEME \cite{memeweb}). 

\vspace{3mm}
\noindent \textbf{\textit{A simple formula for computing the p-value:}} For a motif with a given PM $f$, the p-value is defined as the probability of observing stronger motifs assuming that the sequences are generated according to a background distribution. 

There are different approaches for accurately computing the p-value \cite{stormo, hertzstormo, zhang, yakir, regnier}. While these approaches provide sophisticated methods that precisely compute the p-value, they tend to be complicated to implement. Here, however, as a side-product of our main results, we provide a simple equation that conservatively approximates the p-value.

Specifically, given n sequences, the p-value of a motif $f$ with width $W$ is no more than:
\begin{equation}
pval \approx (n+1)^{W(|{\cal A}|-1)}  2^{-nD(f,g)}
\end{equation} Please see the Appendix A for the detail of derivation of this equation. 

\vspace{5mm}
\noindent \textbf{Methods}\\
\noindent \textbf{\textit{Motif finding problem:}} The motif-finding problem considered here assumed the ``one-occurrence-per-sequence'' model. It is assumed that there are $n$ sequences of length $L$ in the data set (see Fig. 1 for the definition of different parameters). The motifs are assumed to have $W$ columns with a probability matrix denoted by $f$. The motifÕs PM represents the relative frequency of symbols (e.g. DNA bases) in each column of the motif. We measure the strength of a motif by the divergence of its PM from a uniform background distribution $g$. We use the Kullback-Leilber (KL) divergence, also referred to as biological information content \cite{information, stormo}, denoted by $D(f,g) = Iseq (f,g)$ (see Eq. 1).

\vspace{3mm}
\noindent \textbf{\textit{Correction of information content bias due to the sampling error:}} The theoretical result in Eq. 2 is accurate for relatively large $n$. However, in practical application, where the number of sequences is relatively small, e.g. $n<15$ for DNA sequences, a sampling error in computing $f$ causes a bias in the information content $D(f,g)$. We account for this bias by subtracting an approximate term suggested in \cite{information} from the information content used in Eq. 2 as follows:
\begin{equation}
D_{corrected}(f,g)\approx D(f,g)=\frac{|A|-1}{2n\ln(2)}W
\end{equation} where $ln$ is the natural logarithm. The contribution of sampling error vanishes as $n$ increases.

\vspace{3mm}
\noindent \textbf{\textit{Simulations:}} In each experiment, we generated a set of $n$ sequences with length $L$ drawn from a uniform background distribution $g = [0.25~~0.25~~0.25~~0.25]$. We then ran the MEME using OOPS model (only one motif per sequence) and restricted MEME to generate only one motif (the most significant) with width $W$. We repeated the experiment for different number of sequences ($n = \{10, 20, 30\}$), different motif width ($W = \{5, 10, 15\}$), and different sequence lengths ($L = \{50, 100, 500, 1000, 5000\}$). We repeated each experiment for $50$ Monte-Carlo runs resulting in $50$ data points for each experiment. 

For each detected motif, we computed the information content or divergence, $D(f, g)$, using the PMs reported by MEME. Since the input to MEME is a set of random sequences, all detected motifs are supposed to be false-positives. We then compared the false-positives detected by MEME with the theoretical predictions. Each motif detected by MEME is depicted on figures by a star (*). 

\vspace{5mm}
\noindent \textbf{Acknowledgment}\\
The first author would like to acknowledge useful discussions and comments by Alex Nguyen Ba that enhanced the presented results as well as the manuscript significantly. This research is supported by Canadian Institute for Health Research grant $\#202372$ and an infrastructure grant from the Canadian Foundation for Innovation to AMM. 

\vspace{5mm}
\noindent \textbf{Figure legends:}\\
\noindent \textbf{Figure 1. DNA motif finding problem parameters.} In this example, $n=5$ sequences of length $L=80$ are used to detect a motif of width $W=15$. Corresponding probability matrix, $f$, is also shown that represents the relative frequency of nucleotides in each column of the motif. Note that each sequence has only one occurrence of the motif (hence one-occurrence-per-sequence (OOPS) model) 

\vspace{2mm}
\noindent \textbf{Figure 2. Theoretical results compared to MEME simulations.} Theoretical prediction of expected sequence length, $L$, to observe false-positive motifs with information content $D(f,g)$ (solid lines) compared to experimental results of MEME (stars ) for three different motif width $W=5$, $10$, and $15$. The results are for three different number of sequences, $n=\{10,20,30\}$, in the dataset. Each set of experiments are repeated for $50$ Monte-Carlo runs (so there are $50$ stars (*) for each set of experiments). The range of information content is chosen between $0$ and $40$ bits corresponding to what we found for motifs in the Jaspar database \cite{jaspar} (See Supplementary Fig. 6 that shows the frequency of motifs with respect to their corresponding information content). For any given $n$, decreasing $L$ reduces the strength of false-positive motifs. Alternatively, for a fixed $L$, adding more sequences (increasing $n$) reduces the false-positive strength. The dependency of motif strength on $n$ is stronger compared to the dependency on $L$. For instance, that for motifs with $W=10$ in (b), a threefold increase of $n$ (while keeping $L$ constant) reduces $D(f,g)$ by the same amount if $L$ is increased by $2$ orders of magnitudes (while keeping $n$ unchanged). 

\vspace{2mm}
\noindent \textbf{Figure 3. False-positive information content versus the number of sequences.} The dependency on false-positives strengths diminishes with increasing n and reaches a plateau suggesting that it is not necessary to use too many sequences to maintain an acceptable level of false-positives. In this figures, the sequence length is fixed to $L=1000$. Simulation results from MEME, shown by blue (*). There are $50$ simulation results ($50$ stars) for each value of $n$. Simulations are done for $n=\{10, 20, 30, 50, 100\}$. 

\vspace{2mm}
\noindent \textbf{Figure 4. False-positive information content versus the motif width.} False-positive motifsÕ information content, $D(f,g)$, is shown with respect to the motif width for a fixed $L$ and $n$. For the range of motif widths of our interest ($5$ to $20$), the information content is approximately linear in $W$. Given a motif-strength of interest, detecting real motifs with smaller width is easier and less prone to false-positives (i.e. for a given motif strength, shorted motifs rarer). In this figure, the sequence length and the number of sequences are fixed to $L=1000$ and $n=30$, respectively. The theoretical predictions are shown by solid line. The experimental results from MEME are shown by (*). There are $50$ repeated results for each $W=\{5, 10, 15\}$. 

\vspace{2mm}
\noindent \textbf{Figure 5. Examples of applications.} Two real motifs are used to show the application of the theoretical predictions (here motif width is $W=15$). Motifs as strong as ZFP423 \cite{jaspar} in $n=10$ sequences of length $L=1000$ will be buried in false-positives. Therefore, in order to avoid such false-positive motifs, one can reduce $L$ (along Arrow-2) or preferably add more sequences (along Arrow-1) to the dataset. Similarly, it would be very difficult to identify a motif such as the TATA-box motif in a set of $20$ sequences with length $L=100$ due to false-positives. Since using shorter sequences is unlikely, one can increase the number of sequences to $n=30$ (along Arrow-3) to avoid false-positives that have the same strength as the TATA-box. It is interesting to know how strong the false-positive motifs are for motifs with information content equal to the TATA-box but with a width $W=5$ (this is equivalent to trimming all but the core bases of the TATA-box). Fig. 4 shows that this is equivalent to moving along the theoretical curve from $W=15$ to $W=5$ which reduces the false-positive strength enough to detect this motif. 

\begin{figure}[!ht]
    \centering
     {\includegraphics[height = 5in]{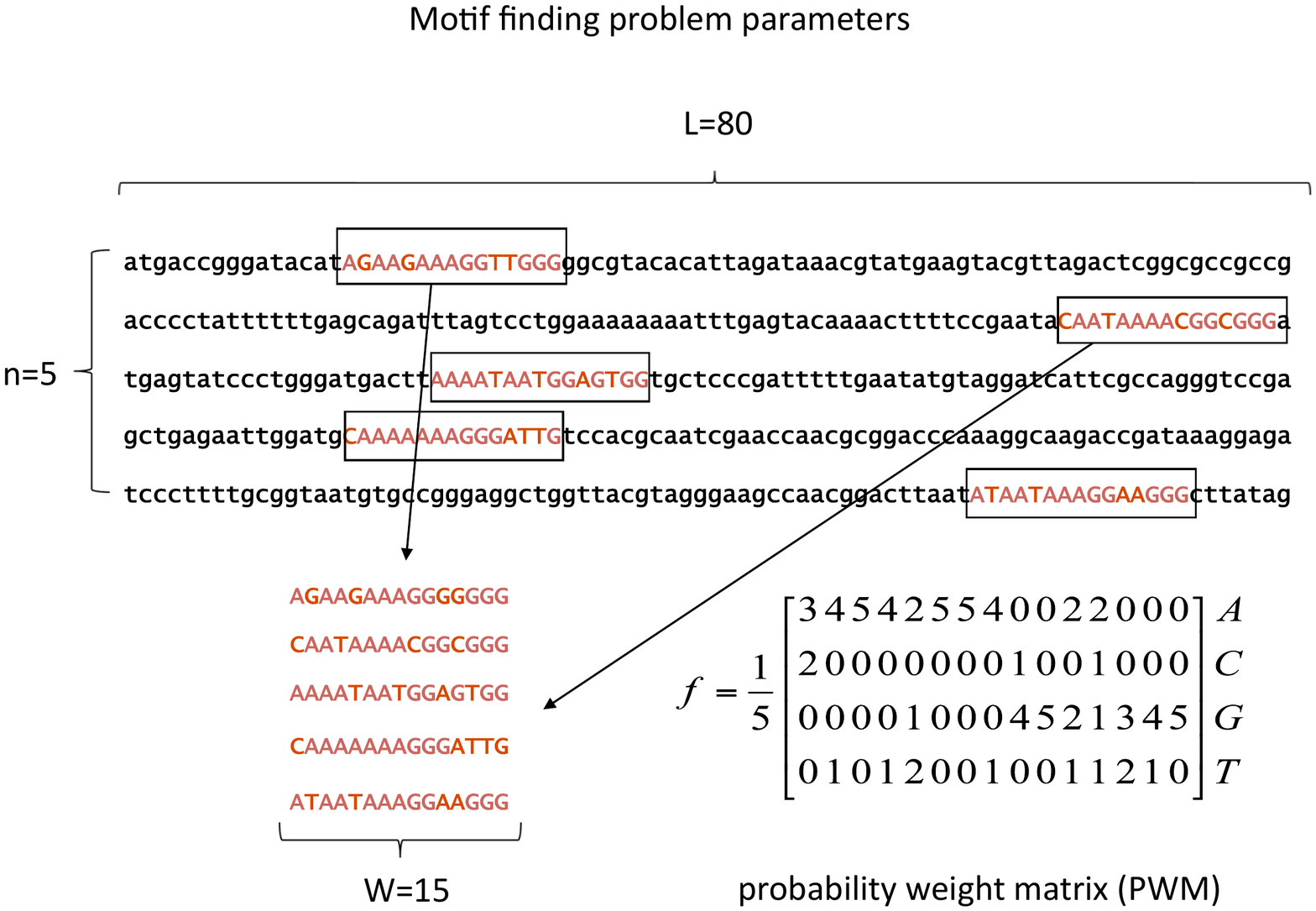}}
\caption{}
\end{figure}

\begin{figure}[!ht]
    \centering
     {\includegraphics[height = 5.5in]{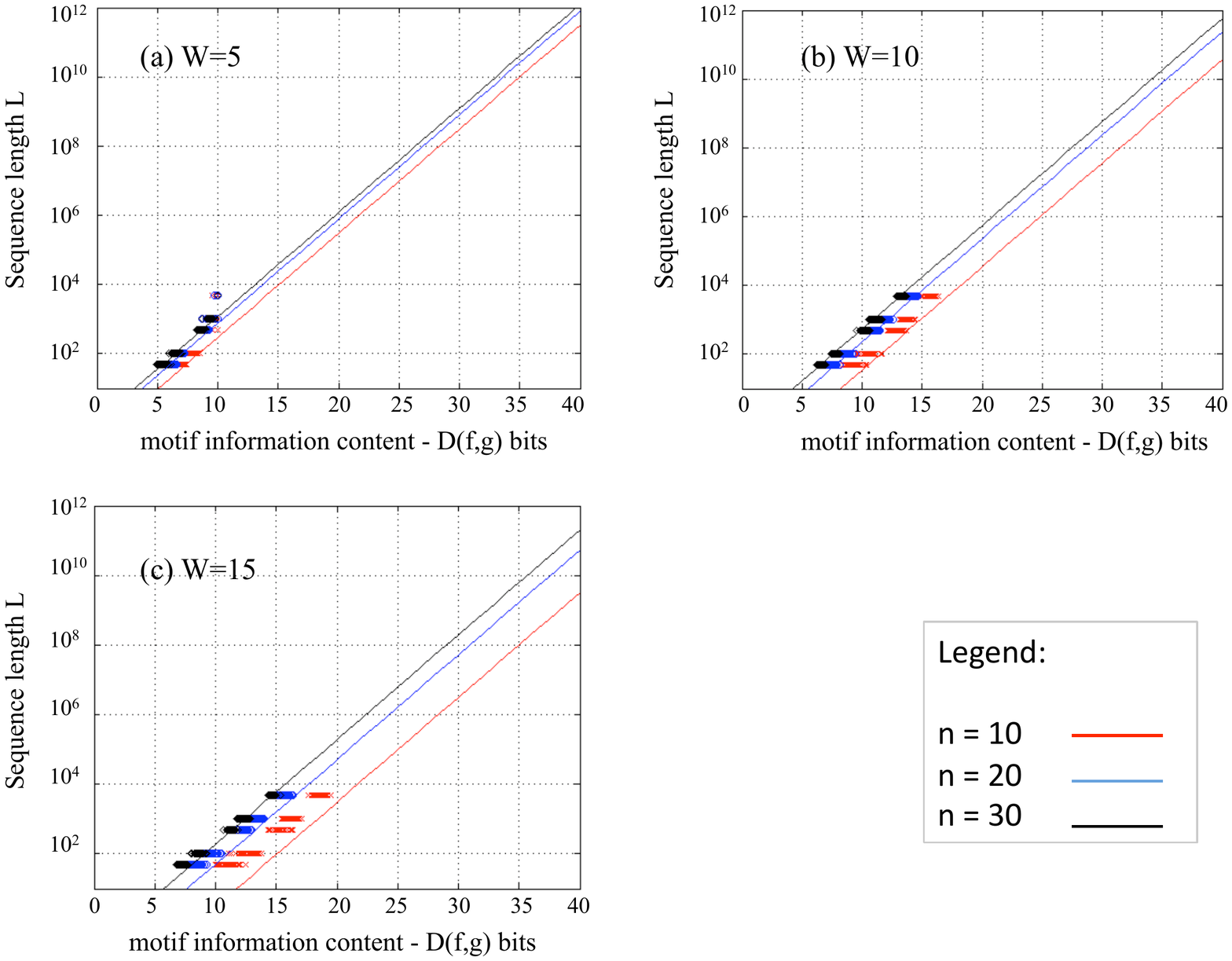}}
\caption{}
\end{figure}

\begin{figure}[!ht]
    \centering
     {\includegraphics[height = 5in]{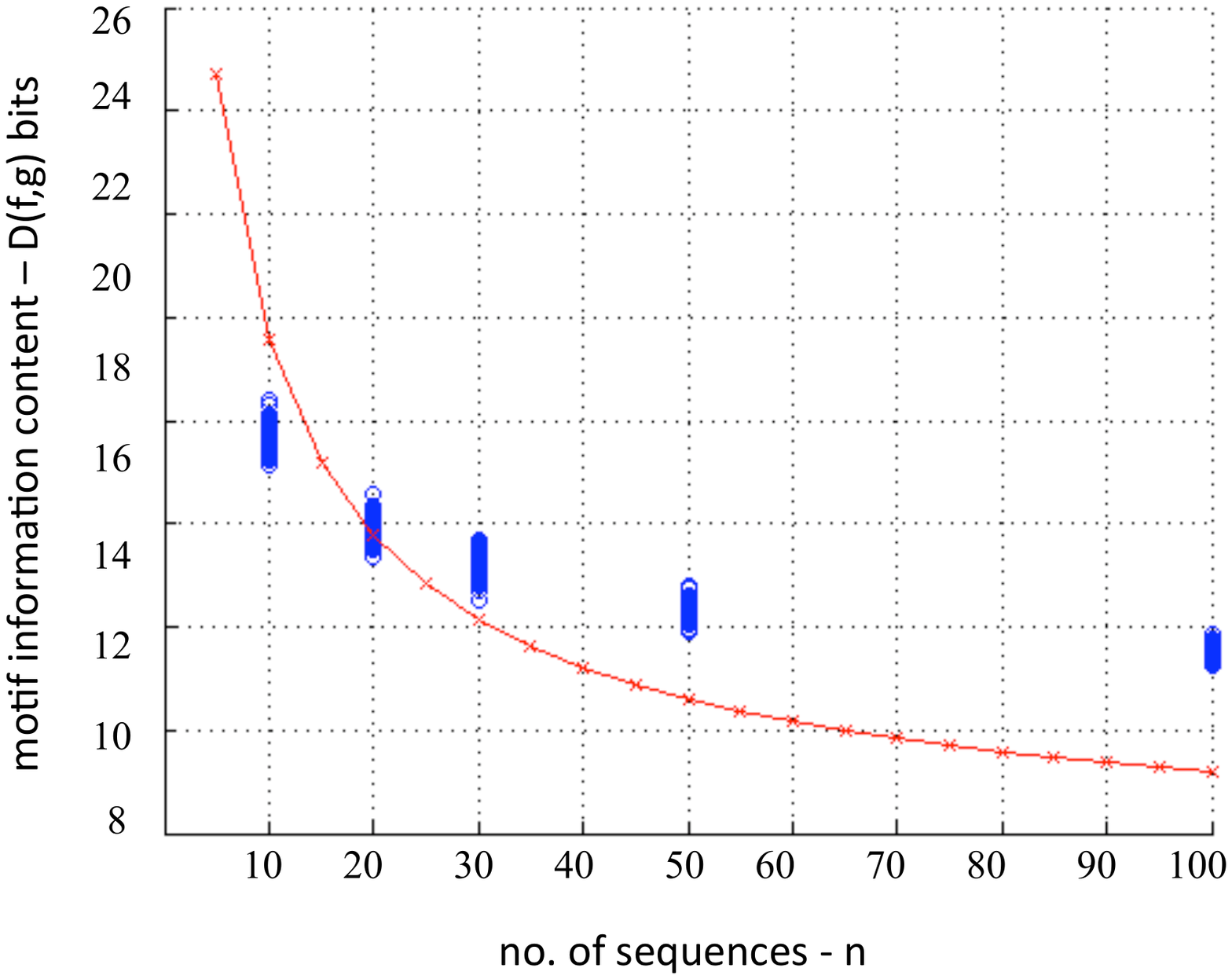}}
   \caption{}
\end{figure}
 
\begin{figure}[!ht]
    \centering
     {\includegraphics[height = 5in]{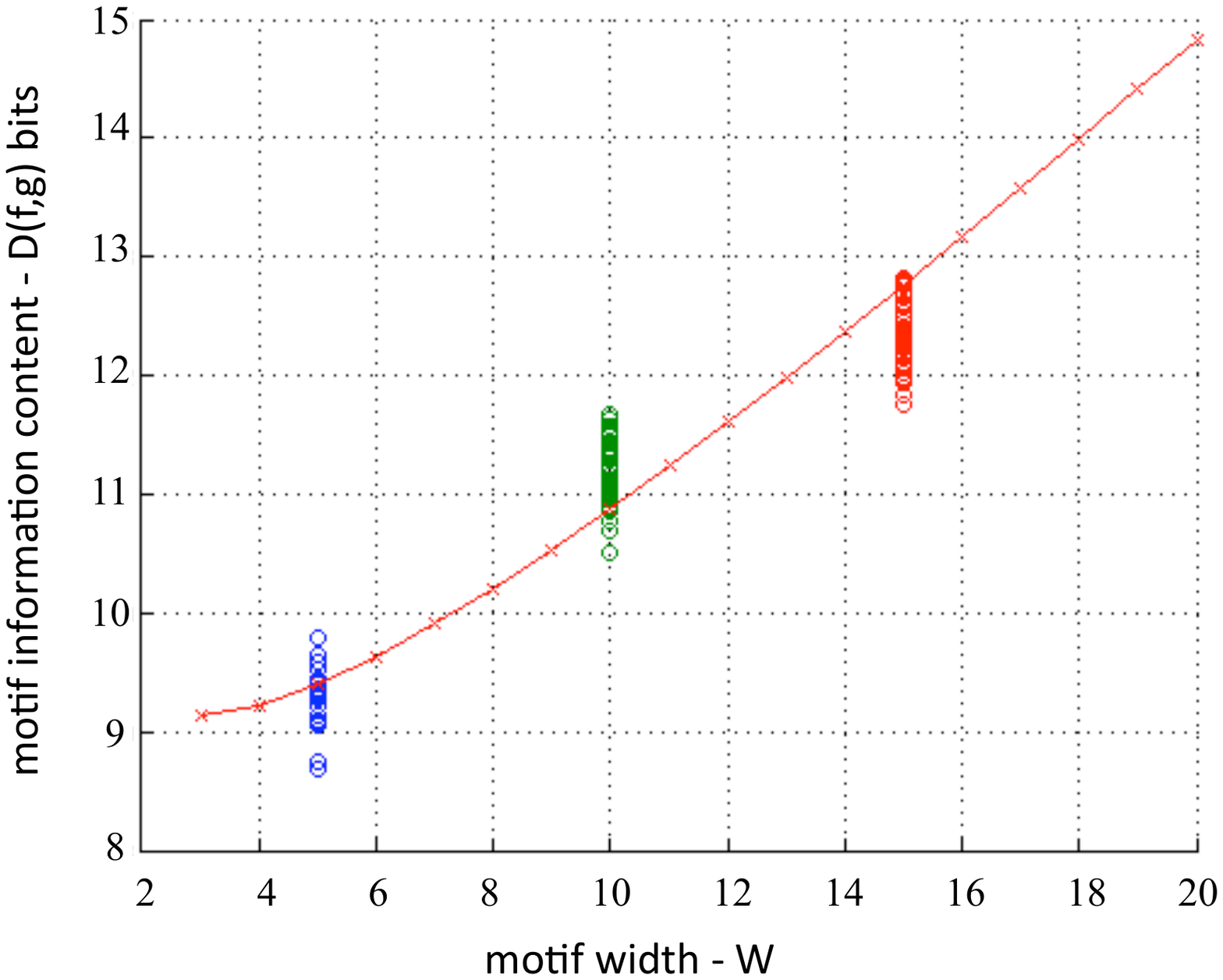}}
\caption{}
\end{figure}
 
\begin{figure}[!ht]
    \centering
     {\includegraphics[height = 5.5in]{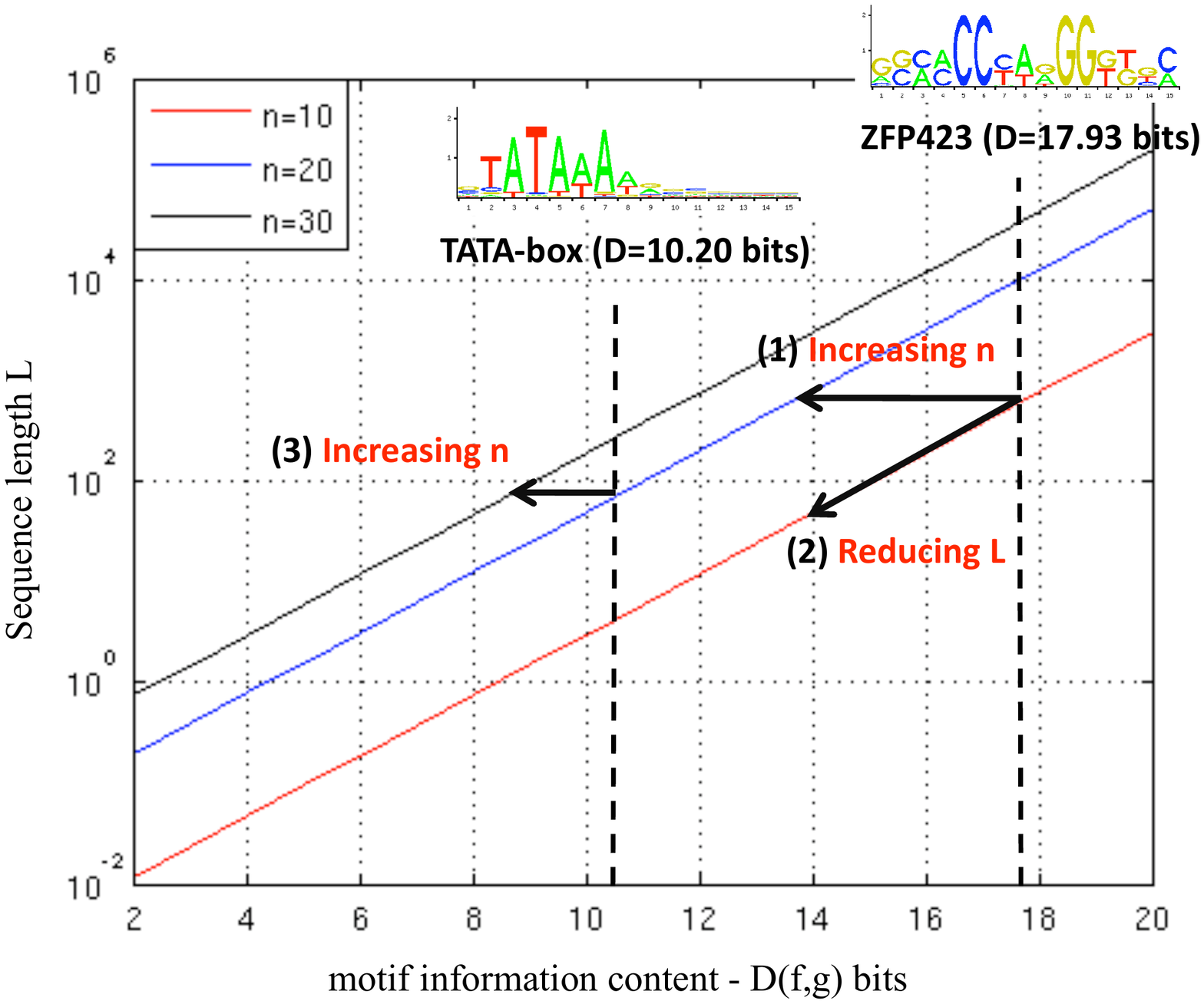}}
\caption{}
\end{figure}
\clearpage
\begin{center}
\textbf{Supplementary Information}
\end{center}
\section{Supplementary figure}
\begin{figure}
    \centering
     {\includegraphics[height = 3in]{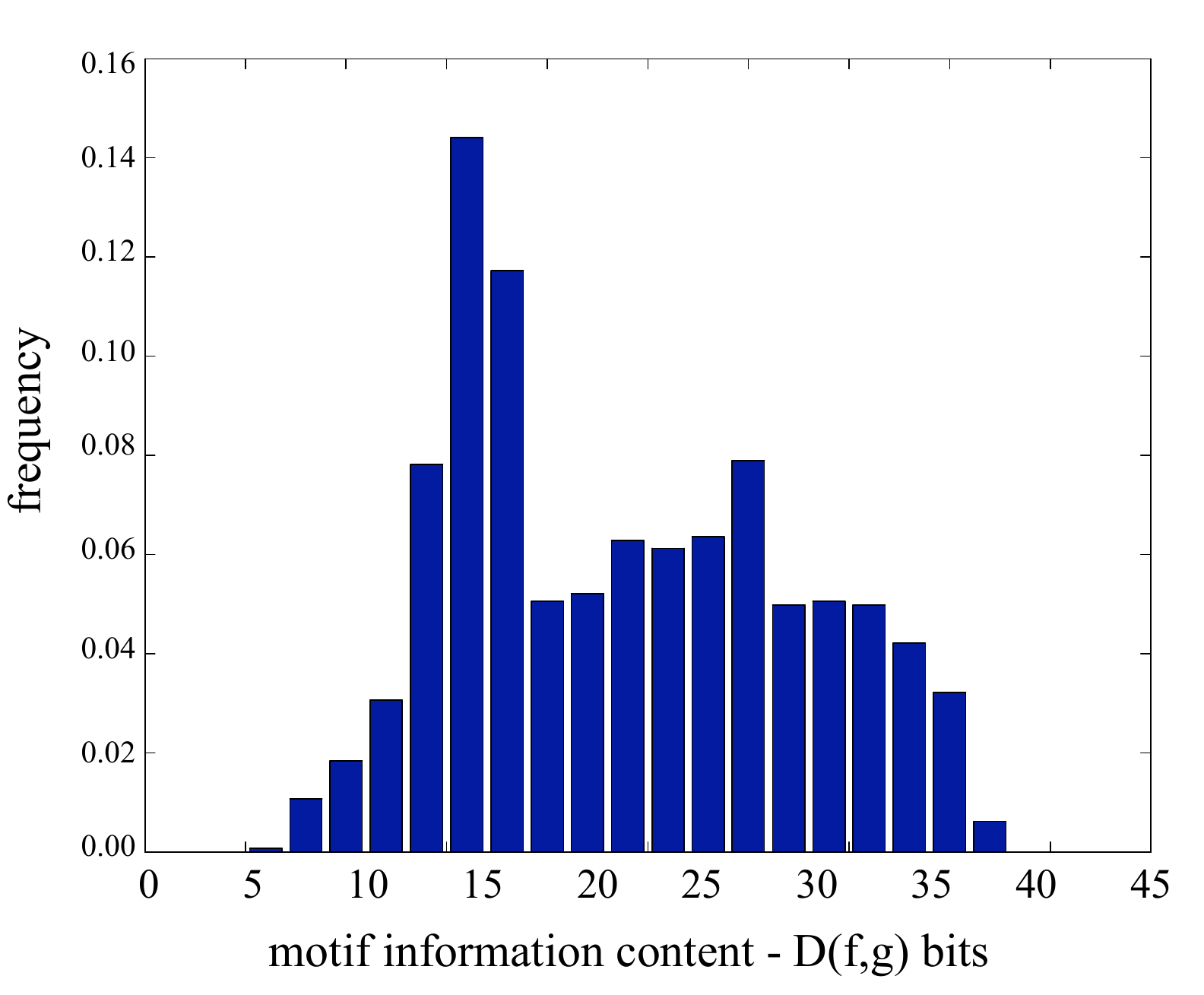}}
\caption{Distribution of motifs in Jaspar database \cite{jaspar} with respect to their information content, $D(f,g)$. The information content is computed using the probability matrix of motifs provided by the database, denoted by f, and assuming a uniform background distribution of nucleotides, $g$. Graphs in Fig. 2 are prepared such that they cover the range of information content of motifs found in this database.}
     \label{fig:histDfg}
\end{figure}

\appendix{Proof of Theorems}
\label{proofs}
Here we provide a series of definitions and lemmas that will be used to prove the main theorem. The proofs for the lemmas used to prove the main theorem are adopted mainly from \cite{cover} with minor changes to apply to the motif-finding problem. The outline of the proof is as follows:
\begin{itemize}
\item We assume that a set of $n$ sequences with length $L$ is generated by a background (nucleotide) distribution $g$.
\item Using a sliding window of width $W$ we form the motif dataset. 
\item We define a divergence function that measures the strength of motifs using their probability matrix (PM).
\item We then compute the probability of observing motifs with a given PM.
\item By adding the probabilities of all motifs with stronger PM than the given motif we compute the p-value of the motif. 
\item We then use the p-value to derive the expected size of the dataset and prove the main theorem.
\end{itemize}
\section{Priliminaries}
Let denote by $Y=[Y_1,~Y_2, ~...~, Y_n]^T$ the set of $n$ sequences with length $L$ used in a typical motif-finding problem ($T$ is the \emph{transpose} of a matrix):
\begin{equation} 
\label{Y}Y = \left[
\begin{array}{cccccc}
 y_{11} &  y_{12} & y_{13} & ... & y_{1(L-1)} & y_{1L}  \\
 y_{21} &  y_{22} & y_{23} & ... & y_{2(L-1)} & y_{2L}  \\
 ... &  ... & ... & ...& ...& ...  \\
 y_{n1} &  y_{n2} & y_{n3} & ... & y_{n(L-1)} & y_{nL}  \\
\end{array}
\right] 
\end{equation} Note that each $Y_i$ is a row vector of L DNA bases or amino-acid residues. In presenting our analysis we consider DNA sequences with alphabets ${\cal A}=\{A,T,G,C\}$. The alphabet size, denoted by $|{\cal A}|$ in this case is equal to $|{\cal A}|=4$. However, the theoretical results are directly applicable to any other alphabet size, including $|{\cal A}|=20$ for protein sequences. 

Motif finding algorithms seek to find a set of over or under-represented short subsequences in $Y$. To prepare the dataset for motif finding, we slide a window of length $W$ on each $Y_i$ shifting by one base at a time to obtain $(L-W+1)$ subsequences of length $W$. We then arrange $n$ number of such subsequences, one for each $Y_i$, to form a motif $X$ as in the following:
\begin{equation} 
\label{X} X = \left[
\begin{array}{cccccc}
 x_{11} &  x_{12} & x_{13} & ... & x_{1(W-1)} & x_{1W}  \\
 x_{21} &  x_{22} & x_{23} & ... & x_{2(W-1)} & x_{2W}  \\
 ... &  ... & ... & ...& ...& ...  \\
 x_{n1} &  x_{n2} & x_{n3} & ... & x_{n(W-1)} & x_{nW}  \\
\end{array}
\right]
\end{equation} Each $X$ is a potential motif. This arrangement is based on the ``one-occurrence-per-sequence'' model in motif finding where each sequence $Y_i$ contributes one and only one subsequence to motif. 

We denote by ${\cal X}$ the set of all motifs $X$. The size of this set is equal to $|{\cal X}|=(L-W+1)^n$.

The search for statistically significant motifs, in essence, involves finding $X \in {\cal X}$ that is distributed differently from a background distribution $g$ (e.g. the distribution of DNA bases genome-wide that is commonly considered to be Uniform). To do so, we represent the motif $X$ by a probability matrix $f$ defined as:
\begin{equation} 
\label{motif}f(X) \triangleq \left[
\begin{array}{cccccc}
 f_{1T} &  f_{2T} & f_{3T} &... & f_{W-1,T} & f_{WT}  \\
 f_{1C} &  f_{2C} &  f_{3C} &... & f_{W-1,C} & f_{WC}  \\
 f_{1A} &  f_{2A} & f_{3A} &... & f_{W-1,A} & f_{WA}  \\
 f_{1G} &  f_{2G} & f_{3G} & ... & f_{W-1,G} & f_{WG}  \\
\end{array}
\right]
\end{equation} where, e.g. $f_{jT}$ denotes the relative frequency of the symbol $T$ in the $j^{th}$ column of the sub-alignment $X$. The PM $f$ represents the empirical distribution of DNA bases at each column of $X$. For sequences of different alphabet, e.g. protein sequences, the PM is defined with $20$ rows corresponding to the number of amino-acid residues. 

Motifs represent the abundance of a particular set of similarly composed short sequences in the set $Y$; a property that is commonly associated with biological importance \cite{stormo}\cite{tompa}\cite{moses}. To quantify the biological importance of a motif we use \emph{information content} measure \cite{information}\cite{moses} that is defined as the divergence of the PM of a motif from a background distribution. Specifically, for a motif with a PM $f$, the divergence from a background distribution $g$ is defined as the Kullback-Leibler (K-L) distance of $f$ and $g$ \cite{stormo}\cite{moses} as in the following:
\begin{equation*}
\label{Dfg}D(f,g) = I_{seq}(f,g) \triangleq \sum_{j=1}^W \sum_{k \in \{T,C,A,G\}} f_{jk} \log \frac{f_{jk}}{g_k}
\end{equation*} where $f_{jk}$ is defined in (\ref{motif}) and $g_k$ is the background distribution of base $k$. 

The divergence, $D(f,g)$, also known as \emph{biological information content} of the motif \cite{information}, is in fact the expected likelihood ratio of the motif given a background distribution $g$ \cite{stormo}. Throughout the manuscript, we refer to a motif $X$ by its PM $f$. We also use the strength of a motif and its information content, interchangeably for the divergence, $D(f,g)$ or $I_{seq}$. 

\section{Probability of a motif}
The probability of a motif $X$, under the background distribution $g$ can be written in terms of its PM using the following lemma \footnote{The PM is the empirical distribution of the motif $X$. In information theory, the empirical distribution is commonly referred to as the \emph{type} of $X$. The discussion presented here is part of the \emph{Method of Types} \cite{types} that studies statistical properties of sequences based on their types.}:
\begin{lem}\label{lem10} If a motif $X$ is drawn \emph{i.i.d} according to $g$, the probability of $X$ under $f$, denoted by $P_g$ throughout the manuscript, depends only on its PM $f$ and is given by: 
\begin{equation}
\label{pgX}P_g(X)=2^{-n(H(f)+D(f,g))}
\end{equation} where H(f) is the binary entropy of $f$ defined as follows:
\begin{equation*}
\label{Hf}H(f) = \sum_{j=1}^W \sum_{k \in \{T,C,A,G\}} f_{jk} \log f_{jk}
\end{equation*} and $D(f,g)$ is defined in (\ref{Dfg}).
\end{lem}

\textit{proof:} See (\cite{cover }, page 281).
\begin{flushright}
$\square$
\end{flushright}

One can compute the probability of a motif $X$ using this lemma. However, in order to compute the probability of observing all motifs that have a PM $f$ we need to add the probabilities of all such motifs as in the following.

\section{Class of a probability matrix and its probability}
Let us define the set of all $X$'s that have the same PM $f$, commonly referred to as \emph{the class} of the PM $f$, as follows:
\begin{equation}
\label{Tf}T(f) \triangleq \{X\in{\cal X} | pm(X)=f\},
\end{equation} If we count the number of motifs in this class and add their corresponding probabilities using (\ref{pgX}) we can compute the probability of all $X$'s with PM $f$. For this purpose, we use the following lemma that gives the size of the class of a PM $f$:
\begin{lem}\label{lem11} The size of the type class of $f$ is upper-bounded as follows: 
\begin{equation}
\label{Tfsize}|T(f)| \leq 2^{nH(f)}
\end{equation}
\end{lem}

\textit{proof:} See (\cite{cover }, page 282).
\begin{flushright}
$\square$
\end{flushright} 

It can be seen from (\ref{Tfsize}) that as $f$ changes in such a way that has a larger entropy (e.g. as it gets closer to a uniform distribution as the background distribution $g$), the total number of motifs, $X$ with PM $f$ becomes exponentially large. Alternatively, when $f$ is such that its entropy is lower, i.e. it is a highly skewed PM, the number of motifs with a PM equal to $f$ becomes exponentially small. 

Now, to compute the probability of observing motifs with a PM $f$, one can add up the probabilities of all $X\in T(f)$ as follows:
\begin{lem}\label{lem12} If motifs $X$ are drawn \emph{i.i.d} according to a distribution $g$, the probability of observing motifs that all have a PM $f$ is upper-bounded as follows:
\begin{equation}
\label{pgTf}P_g(T(f)) \leq 2^{-nD(f,g)}
\end{equation}
\end{lem}

\textit{proof:} The probability of a class $T(f)$ can be written as:
\begin{eqnarray}
\nonumber P_g(T(f)) &=& \nonumber \sum_{X\in T(f)} P_g(X) 
\\ \label{inq1} &=& \sum_{X \in T(f)} 2^{-n(D(f,g)+H(f))} 
\\ \nonumber &=& |T(f)| 2^{-n(D(f,g)+H(f))} 
\\ \label{inq3} &\leq& 2^{nH(f)} 2^{-n(D(f,g)+H(f))} 
\\ \nonumber &=& 2^{-nD(f,g)}
\end{eqnarray} where in (\ref{inq1}) we used (\ref{pgX}) of Lemma \ref{lem10} and in (\ref{inq3}) we used (\ref{Tfsize}) of Lemma \ref{lem11}. 
\begin{flushright}
$\square$
\end{flushright}

According to this lemma, the probability of observing motifs with a PM $f$ is exponentially proportional to the distance of $f$ and $g$. Therefore, as $f$ gets closer (in the KL divergence sense) to $g$, the probability of observing motifs becomes closer to $1$. On the other hand, the probability of strong motifs with large divergence from background, i.e. larger $D(f,g)$, is exponentially small.

We now can compute the probability of motifs with PM $f$. By adding up the probabilities of motifs with stronger PMs we can compute the p-value of a motif as in the following. Before that, we need to count all possible PMs:

\section{Number of possible PMs}
Enumerating all PM is impractical for larger $n$. We, instead in the following, drive a bound on the number of possible PMs.

First, let us consider only one column of a motif $X$ in (\ref{X}) with a PM as in (\ref{motif}). There are $n$ sequences in the motif. Therefore, the column has $n$ symbols chosen from the alphabets in ${\cal A}$. One can enumerate all possible distributions of bases in this column:
\begin{equation*}
{\cal P}=\bigg\{\big(f_{1}^T,  f_{1}^C,  f_{1}^A,  f_{1}^G\big): (\frac{0}{n},\frac{0}{n}, \frac{0}{n}, \frac{n}{n}\big), \big(\frac{0}{n},\frac{0}{n}, \frac{1}{n}, \frac{n-1}{n}\big), ..., \big(\frac{n}{n},\frac{0}{n}, \frac{0}{n}, \frac{0}{n}\big)\bigg\}
\end{equation*} It can be seen that the numerator of frequencies change from $0$ to $n$. Furthermore, there are three independent frequencies in this PM, i.e. the last one is fixed by the rest to have a sum equal to $1$. Therefore, there are about $(n+1)^3$ different possible arrangements of this frequencies. We formalize this idea for an extended number of columns, $W$, in the following lemma \cite{cover}:
\begin{lem}\label{lem1} For a motif of width $W$, there are at most  $|{\cal P}| \leq (n+1)^{W(|{\cal A}|-1)}$ PMs in ${\cal P}$.
\end{lem}

\textit{proof:} There are $|{\cal A}|-1$ components in the PM of any column (the last component is fixed by the the others). The numerator of each component can take $n+1$ values. Therefore, each column can have $(n+1)^{|{\cal A}|-1}$ PMs. Since each column is independently and identically distributed, there are $(n+1)^{W(|{\cal A}|-1)}$ different PMs for the motif of width $W$. 
\begin{flushright}
$\square$
\end{flushright}

\section{An approximate value for p-value}
By defining the maximum number of PMs for a motif of width $W$ and knowing the probability of the class of each PM (Lemma \ref{lem12}) we can now compute the p-value of a motif with PM $f$. The main idea, as explained before, is to first define the set of all motifs with PMs stronger than $f$, i.e. with $D\geq D(f,g)$ and then to use Lemma \ref{lem12} to compute its probability. This idea is formalized in the following theorem, known as Sanov's Theorem. Here we provide a simplified version of the proof that is only applicable to our case. Interested readers are referred to (\cite{cover}, page 292) for general theorem and technical details.

\begin{lem} \label{lem2} Given that a set ${\cal X}$ is generated according to a background distribution, $g$, the probability of observing motifs $X$ with PMs that are diverged from the background at least by $D(f,g)$ is upper bounded by:
\begin{equation}
\label{Pg}P_g(X) \leq (n+1)^{W(|{\cal A}|-1)}  2^{-nD(f,g)}
\end{equation} where $P_g$ is the probability under the background distribution $g$. 
\end{lem}

\textit{proof:} We denote by ${\cal E}(f)$ the set of all motifs, $X$, that have a PM $h$ that is diverged from $g$ at least by $D(f,g)$: 
\begin{equation}
\label{calE}{\cal E}(f) \triangleq \{X\in{\cal X} | pm(X)=h, D(h,g)\geq D(f,g)\},
\end{equation} By definition, the probability of the set ${\cal E}$ is the p-value of motif with a PM $f$. The probability of the set ${\cal E}$ is equal to the sum of the probabilities of the classes of PMs in ${\cal E}$. We have:
\begin{eqnarray}
\label{q0} P_g({\cal E})  &=& \sum_{h \in {\cal E}}P_g(T(h)) \\ \label{q1} &\leq& \sum_{h \in {\cal E}} 2^{-nD(f,g)} \\ \label{q2} &\leq& \sum_{h \in {\cal E}} \max_{h \in {\cal E}} 2^{-nD(h,g)} \\ \label{q3} &=&  \sum_{h \in {\cal E}} 2^{-n \min_{h \in {\cal E}} D(h,g)} \\ \label{q4} &\leq& \sum_{h \in {\cal E}} 2^{-n D(f,g)} \\ \label{q5} &=& 2^{-n D(f,g)} \sum_{h \in {\cal E}}(1) \\ \label{q6} &\leq& 2^{-n D(f,g)} (n+1)^{W(|{\cal A}|-1)}
\end{eqnarray} In (\ref{q0}) we used the fact that, by definition, the probability of the set ${\cal E}$ is the sum of probabilities of the classes of PMs in {\cal E}. In (\ref{q1}) we used (\ref{pgTf}) of Lemma \ref{lem12} that gives an upper-bound on the probability of the class of a PM $h$. Inequality (\ref{q2}) in valid in we replace all $2^{-nD(h,g)}$ in summation with its maximum value. Similarly, this is valid if we replace its exponent with its minimum in Inequality (\ref{q3}). By definition of the set ${\cal E}$ in (\ref{calE}), all its PMs, i.e. all $h\in{\cal E}$ have a divergence not less than $D(f,g)$. Therefore, we have $\min_{h \in {\cal E}} D(h,g) = D(f,g)$ in Inequality (\ref{q4}). It can be seen in (\ref{q5}) that $D(f,g)$ is independent of the summation and therefore can be taken out. In (\ref{q6}) we replace the summation with the number of its components, defined by the total number of possible PMs given by Lemma \ref{lem1}.
\begin{flushright}
$\square$
\end{flushright}

This Lemma provides an approximate equation (in fact an upper-bound) for the p-value of a motif with PM $f$ presented in Eq. 2.

\section{Proof of the main theorem (Eq. 2)}
\begin{thm} \label{thm1}Given a set ${\cal Y}$ of $n$ sequences of symbols from an alphabet $|{\cal A}|$, the expected sequence length, $L$, in order to observe at least one motif of width $W$ and with a PM diverged at least as much as $D(f,g)$ is given by:
\begin{equation}
\label{Lf}L \approx \frac{W2^{D(f, g)}}{\big((n+1)^{W(|{\cal A}|-1)}\big)^{1/n}}
\end{equation} where $|{\cal A}|$ is the cardinality of the set ${\cal A}$, e.g. $|{\cal A}|=4$ for DNA sequences with ${\cal A}=\{A,T,C,G\}$. This is an approximate lower bound on the expected length. 
\end{thm}
\begin{proof}

The Lemma \ref{lem2} gives an upper-bound on the probability of observing motifs with a type that is diverged greater than $D(f,g)$. This probability when multiplied with the total number of motifs, $X$, in the set ${\cal X}$, gives an upper-bound on the number of such motifs observed, as derived in the following. 

Note that the total number of $X$'s in the data set ${\cal X}$ is equal to $|{\cal X}| = (L-W+1)^n$. However, it can be easily verified that for large $L$, each $X$ is overlapped with at least $2$ neighboring  $X$'s due to a one-shift-at-a-time sliding window. This results in an approximately $|{\cal X}| \approx (L-W+1/W)^n$ effectively independent $X \in {\cal X}$. Therefore, the expected number of observations of $X \in {\cal E}(f)$, denoted by $N_f$ is approximately:
\begin{eqnarray}
\nonumber N_{f} &\approx&  \nonumber |{\cal X}| P_g({\cal E}(f)) \\ &\leq& \label{Nf} (L-W+1/W)^n (n+1)^{W(|{\cal A}|-1)} 2^{-nD(f, g)}
\end{eqnarray} This is in fact an upper-bound on the number of expected motifs observed. 

By letting $1 \leq N_{f}$, the minimum expected length to observe at least one $X$ with $pmw(X)=f$, becomes:
\begin{eqnarray}
\nonumber L &\geq& \nonumber W-1+\left(\frac{W^n2^{nD(f, g)}}{(n+1)^{W(|{\cal A}|-1)}}\right)^{1/n} \\ &\approx& \label{Lfproof} \frac{W2^{D(f, g)}}{(n+1)^{W(|{\cal A}|-1)/n}} 
\end{eqnarray} where we used the fact that in motif-finding problems we have $L\gg W-1$. 
\end{proof}

\appendix{Extension to other motif-finding models}

The proposed method here assumes the ``one-occurrence-per-sequence'' model in motif finding (similar to the OOPS model in MEME \cite{meme}). However, the analysis is extendable to other models by appropriately redefining the space of all motifs in the dataset. For instance, in cases where each sequence can carry either zero or one motif (similar to ZOOPS model in MEME \cite{meme}), the following equation provides the expected length $L$:
\begin{equation*}
L \approx \frac{W2^{rD(f, g)}}{(rn+1)^{W(|{\cal A}|-1)/n}} 
\end{equation*} where $(r \leq 1)$ is the fraction of sequences that carry a motif (note that this equation simplifies to Eq. 2 for $(r = 1)$). In this equation, the denominator is always larger than $1$. Therefore, the expected length is reduced significantly compared to OOPS model, suggesting a potentially higher rate of false-positives in ZOOPS models.


\begin{thebibliography}{4}
\bibitem{tompa} Tompa M. et al, Assessing computational tools for the discovery of transcription factor binding sites, Nature Biotechnology, Vol. 23, No. 1, pp. 137-144, Jan. 2005.
\bibitem{sandve} Sandve G.K, et al, Improved benchmarks for computational motif discovery, BMC Bioinformatics, 8:193, 2007.
\bibitem{hu} Hu J. et al, Limitations and potentials of current motif discovery algorithms, Nucleic Acids Research, Vol. 33, No. 15, pp. 4899-4913, 2005.
\bibitem{wasserman} Wasserman W.W. and Sandelin A, Applied bioinformatics for the identification of regulatory elements, Nature Reviews Genetics 5, pp. 276-287, April 2004.
\bibitem{das} Das M.K., and Dai H.K., A survey of DNA motif finding algorithms, BMC Bioinformatics, 8 (Suppl 7): S21, 2007.
\bibitem{moses} Moses, A.M., and Sinha, S., Regulatory Motif Analysis, In: D. Edwards et al. (eds.), Bioinformatics: Tools and Applications, pp. 137-163, Springer Science+Business Media LLC (2009).
\bibitem{harbison} Harbison C.T. et al, Transcriptional regulatory code of a eukaryotic genome, NATURE, Vol.431, pp. 99-104, 2 Sept. 2004.
\bibitem{cover} Cover T.M. and Thomas J.A., Elements of information theory, Wiley Interscience, New York, 1991.
\bibitem{information} Schnider T.D. et al., Information content of individual genetic sequences, Journal of Theoretical Biology, Vol. 189, No. 4, pp. 427-441, 1997.
\bibitem{stormo} Stormo G.D., DNA binding sites: representation and discovery, Bioinformatics, Vol. 16, No. 1, pp. 16-23, 2000.
\bibitem{meme} Bailey T. L. et al, Discovering and analyzing DNA and protein sequence motifs, Nucleic Acids Research, Vol. 34, Web Server issue, W369-W373.
\bibitem{memeweb} The MEME Suite: http://meme.sdsc.edu
\bibitem{jaspar} Bryne J.C., et al., JASPAR, the open access database of transcription factor-binding profiles: new content and tools in the 2008 update, Nucleic Acids Research. Jan. 2008 (Database issue).
\bibitem{hertzstormo} Hertz G.Z, and Stormo G.D., Identifying DNA and protein patterns with statistically significant alignments of multiple sequences, Bioinformatics, Vol. 15, No. 7/8, pp. 563-577, 1999.
\bibitem{zhang} Zhang J., et al., Computing exact p-values for DNA motifs, Bioinformatics, Vol. 23 No. 5, pp. 531-537, 2007.
\bibitem{yakir} Siegmund D., and Yakir B., Approximate p-values for local sequence alignments, The Annals of Statistics, Vol. 28, No. 3, pp. 657-680, Jun. 2000.
\bibitem{regnier} Regnier M. and Vandenbogaert W., Comparison of statistical significance criteria, Journal of Bioinformatics and Computational Biology, Vol. 12, No. 12, October 6, 2005.
\bibitem{types} Csiszar I., Method of types, IEEE Trans. on Information Theory, Vol. 44, pp. 2505--2523, Oct. 1998.
\end{thebibliography}
\end{document}